\documentclass[prb,superscriptaddress,showkeys,nobibnotes,showpacs]{revtex4}

\begin{document}

\newtheorem{proposition}{Proposition}[section]
\newtheorem{lemma}{Lemma}[section]
\newtheorem{theorem}{Theorem}[section]
\newtheorem{definition}{Definition}[section]
\newtheorem{corollary}{Corollary}[section]

\newcommand{\nablat}{\tilde{\nabla}}
\newcommand{\signa}{\varepsilon_{A}}
\newcommand{\ds}{\mbox{ds}}
\newcommand{\dt}{\mbox{dt}}
\newcommand{\dx}{\mbox{dx}}
\newcommand{\dy}{\mbox{dy}}
\newcommand{\dz}{\mbox{dz}}


\title{Ideally embedded space-times}

\date{\today}

\author{Stefan \surname{Haesen}}
\email{Stefan.Haesen@kubrussel.ac.be}
\affiliation{Group of Exact Sciences, Katholieke Universiteit Brussel, Vrijheidslaan 17, 1081 Brussel, Belgium}

\author{Leopold \surname{Verstraelen}}
\affiliation{Department of Mathematics, Katholieke Universiteit Leuven, Celestijnenlaan 200 B, 3001 Heverlee, Belgium}

\begin{abstract}
Due to the growing interest in embeddings of space-time in higher-dimensional spaces we consider a specific type of embedding. After proving an inequality between intrinsically defined curvature invariants and the squared mean curvature, we extend the notion of ideal embeddings from Riemannian geometry to the indefinite case. Ideal embeddings are such that the embedded manifold receives the least amount of tension from the surrounding space. Then it is shown that the de Sitter spaces, a Robertson-Walker space-time and some anisotropic perfect fluid metrics can be ideally embedded in a five-dimensional pseudo-Euclidean space.
\end{abstract}



\maketitle


\section{Introduction}

In recent years the ideas of Kaluza and Klein have received new attention. Shortly after the publication of the General Theory of Relativity Kaluza proposed to unify gravity and electromagnetism by adding an extra dimension. Klein suggested that this fifth dimension would be compactified and unobservable on experimentally accessible energy scales. This idea of compactifying the extra dimension has dominated the search for a unified theory and lead to the eleven-dimensional supergravity theory and more recent ten-dimensional superstring theory (see Ref.~\onlinecite{overduin} for an overview).

Instead of compactifying the extra dimensions other approaches have been developed. In the Space-Time-Matter (STM) theory \cite{wesson} the $(3+1)$-dimensional cosmologies may be recovered from the geometry of $(4+1)$-dimensional, vacuum General Relativity. Matter in four dimensions is induced by the shape of the embedded hypersurface and the five-dimensional Ricci flat geometry. More recently the Randall-Sundrum scenario has gained a lot of support. In Ref.~\onlinecite{randall1,randall2} they try to solve the hierarchy problem between the observed Planck and weak scales by embedding the three-brane in a non-factorizable five-dimensional metric.

From a mathematical point of view the theory of embeddings starts with the definition of a manifold by Riemann.  Shortly after the publication of his famous Habilitationsschrift (see e.g. Ref.~\onlinecite{spivak} for a translation) Schl\"{a}fli\cite{eisenhart} conjectured that any n-dimensional Riemannian manifold could be locally and isometrically embedded in a $d$-dimensional Euclidean space with $d=n(n+1)/2$. This was proven by Janet and Cartan and extended to manifolds with indefinite metric by Friedman \cite{friedman}. The Janet-Cartan theorem as it became known implies that we at maximum need ten dimensions to locally and isometrically embed any four-dimensional space-time. 

A lesser known theorem by Campbell and Magaard\cite{lidsey} states that any analytical Riemannian space $V_{n}(s,t)$ can be locally and isometrically embedded in a Ricci flat Riemannian space $V_{n+1}(\tilde{s},\tilde{t})$, with $\tilde{s}=s+1, \tilde{t}=t$ or $\tilde{s}=s, \tilde{t}=t+1$. This theorem has obvious applications in STM-theory \cite{seahra}. For further generalizations of the Campbell-Magaard theorem to embedding spaces which are Einstein, scalar field sourced or have nondegenerate Ricci tensor see Ref.~\onlinecite{anderson,dahia1,dahia2}.

In applications of the embedding theorems one often starts from a given metric and looks for the embedding space with the minimal dimension or one puts restrictions on the source type \cite{collinson,rosen,stephani}. In the following we will take a different approach by putting a restriction on the type of embedding. Using some recently defined intrinsic curvature invariants on a manifold we prove an inequality between intrinsic and extrinsic curvatures of an embedded Lorentzian manifold in a pseudo-Euclidean space. For a proof in the Riemannian case see Ref.~\onlinecite{chen2}. An embedding for which the equality holds is called ideal and in this case the shape operators take on specified forms. The space-times which satisfy such an ideal embedding in a five-dimensional space are determined. In the remainder all embeddings are local and isometric.


\section{$\Lambda$-curvatures of Chen}

Starting from a Lorentzian manifold $(M,g)$ with signature $(m-1(+),1(-))$ isometrically embedded in a pseudo-Euclidean space $(E_{n},\eta)$ of signature $(n-1,1)$ or $(m-1,n-m+1)$ we will introduce the intrinsically defined $\Lambda$-curvature invariants of Chen \cite{chen}. 

{\par\noindent}We denote the Levi-Civita connection on $M$ with $\nabla$ and on $E_{n}$ with $\tilde{\nabla}$. The covariant derivative in $E_{n}$ between two tangent vectors $X$ and $Y$ on $M$ can be decomposed in a tangential and normal part,

\[ \tilde{\nabla}_{X}Y = \nabla_{X}Y +\Omega(X,Y)\ , \]

{\par\noindent}with $\Omega: TM\times TM\rightarrow N(M)$ the second fundamental form. If we choose an orthonormal basis $\{\xi_{A}\}$ in the normal space $N(M)$ of $M$ and denote the signature of the basis vectors with $\signa=\eta(\xi_{A},\xi_{A})=\pm 1$, we can define $\Omega$ as

\begin{equation} 
\Omega(X,Y) = \sum_{A=m+1}^{n}\signa \eta(\nablat_{X}Y, \xi_{A}) \xi_{A}\ . \label{defomega}
\end{equation}

{\par\noindent}In the following Greek indices run from 1 to m, Latin indices from 1 to n and capital indices from m+1 to n, unless otherwise stated.

{\par\noindent}The integrability conditions for the existence of an embedding are given by the Gauss-Codazzi-Ricci equations \cite{goenner1,goenner2},

\begin{eqnarray}
R_{\alpha\beta\gamma\mu} & = & \sum_{A}\signa \{ \Omega_{\alpha\gamma}^{A}\Omega_{\beta\mu}^{A} - \Omega_{\alpha\mu}^{A}\Omega_{\beta\gamma}^{A}\}\ , \label{gauss} \\
\nabla_{\gamma}\Omega_{\alpha\beta}^{A} -\nabla_{\beta}\Omega_{\alpha\gamma}^{A} & = & \sum_{B}\varepsilon_{B} \{ S^{BA}_{\ \ \gamma}\Omega_{\alpha\beta}^{B} - S^{BA}_{\ \ \beta}\Omega_{\alpha\gamma}^{B} \}\ , \label{codazzi} \\
\nabla_{\beta}S^{BA}_{\ \ \alpha} -\nabla_{\alpha}S^{BA}_{\ \ \beta} & = & \sum_{C}\varepsilon_{C}\{ S^{CB}_{\ \ \beta}S^{CA}_{\ \ \alpha}- S^{CB}_{\ \ \alpha}S^{CA}_{\ \ \beta}\}  + g^{\gamma\mu}\{\Omega_{\gamma\beta}^{B}\Omega_{\mu\alpha}^{A} -\Omega_{\gamma\alpha}^{B}\Omega_{\mu\beta}^{A}\}\ , \label{ricci}
\end{eqnarray}

{\par\noindent}with $S^{AB}_{\ \ \alpha}$ the torsion vector. For an interpretation of this vector as a gauge field in a Kaluza-Klein view of embeddings see Ref.~\onlinecite{maia1} and as a real connection on space-time see Ref.~\onlinecite{maia2}.

{\par\noindent}The mean curvature vector is defined as

\[ \vec{H} = \sum_{A}\signa g^{\alpha\beta}\Omega_{\alpha\beta}^{A}\, \xi_{A}\ . \]

{\par\noindent}Let $\{ e_{\alpha}\}$ be an orthonormal basis of $M$. The sectional curvature of a two-plane spanned by the orthonormal vectors $\{e_{\alpha},e_{\beta}\}$ is defined by

\[ K(e_{\alpha}\wedge e_{\beta}) = \varepsilon_{\alpha\beta} g(R(e_{\alpha},e_{\beta})e_{\beta},e_{\alpha})\ , \]

{\par\noindent}with $\varepsilon_{\alpha\beta}=\varepsilon_{\alpha}\varepsilon_{\beta}$. The scalar curvature of an r-plane section $L$ spanned by the orthonormal vectors $\{e_{1},\ldots,e_{r}\}$ is defined as

\[ \tau(L) = \sum_{\alpha<\beta}K(e_{\alpha}\wedge e_{\beta})\ ,\ 1\leq\alpha<\beta\leq r\ . \]

{\par\noindent}The scalar curvature of the whole Lorentzian manifold is denoted by $R$. Denote the constant $c(n_{1},\ldots,n_{k})$ by 

\[ c(n_{1},\ldots,n_{k}) = \frac{2(m+k-\sum_{j=1}^{k}n_{j})}{m+k-1-\sum_{j=1}^{k}n_{j}}\ . \]

{\par\noindent}We are now in a situation to define the $\Lambda$-curvature invariants of Chen in the pseudo-Riemannian case as

\begin{eqnarray*}
\lefteqn{ \Lambda(n_{1},\ldots,n_{k}) = }  \\
 & &
c(n_{1},\ldots,n_{k}) \left[ R -\mbox{inf}\{ \tau(L_{1})+\ldots+\tau(L_{k})\mid L_{j}\, \mbox{a non-null}\, n_{j}-\mbox{plane section},\, L_{i}\perp L_{j}\}\right]\ ,  
\end{eqnarray*}

{\par\noindent}and

\begin{eqnarray*}
\lefteqn{\hat{\Lambda}(n_{1},\ldots,n_{k}) = } \\
 & &  c(n_{1},\ldots,n_{k}) \left[ R -\mbox{sup}\{ \tau(L_{1})+\ldots +\tau(L_{k})\mid L_{j}\, \mbox{a non-null}\, n_{j}-\mbox{plane section},\, L_{i}\perp L_{j}\} \right]\ . 
\end{eqnarray*}

{\par\noindent}Note that in our definition the plane sections can be timelike or spacelike. Let $\{e_{1},\ldots,e_{m},\xi_{m+1},\ldots,\xi_{n}\}$ be an orthonormal basis of $E_{n}$. Because we have space-time applications in mind we take $M$ to be time-orientable, i.e. there exists a global nowhere-zero timelike vector field which we denote with $e_{m}$. From (\ref{defomega}) we have

\[ \Omega_{m\alpha}^{A} = -\eta(e_{m},\tilde{\nabla}_{e_{\alpha}}\xi_{A}) = -\eta(e_{\alpha},\tilde{\nabla}_{e_{m}}\xi_{A})\ , \]

{\par\noindent}with $A=m+1,\ldots,n$ and $\alpha=1,\ldots,m-1$.

\begin{definition}
An embedding $x: (M,g)\rightarrow (E_{n-1,1}, \eta)$ is called causal-type preserving if $\tilde{\nabla}_{e_{\alpha}}\xi_{A}$ is spacelike, $\forall\, A=m+1,\ldots,n$ and $\forall\, \alpha=1,\ldots,m-1$.
\end{definition}

\begin{definition}
An embedding $x: (M,g)\rightarrow (E_{m-1,n-m+1}, \eta)$ is called causal-type preserving if $\tilde{\nabla}_{e_{m}}\xi_{A}$ is timelike, $\forall\, A=m+1,\ldots,n$.
\end{definition}

{\par\noindent}From the above we see that causal-type preserving embeddings have $\Omega_{m\alpha}^{A}=0$, $\forall \alpha=1,\ldots,m-1$.


\section{Ideal embeddings}

{\par\noindent}We can now formulate and proof an inequality relating the above intrinsically defined curvature invariants and the square of the extrinsic mean curvature of the embedded manifold.

\begin{theorem}
Let $x: (M,g)\rightarrow (E_{n},\eta)$ be a causal-type preserving embedding of a Lorentzian m-dimensional manifold in a n-dimensional pseudo-Euclidean manifold. For any k-tuple $(n_{1},\ldots,n_{k})$ we have that

\begin{equation}
\| H\|^{2} \geq \Lambda(n_{1},\ldots,n_{k})\ , \label{ineq1}
\end{equation}

{\par\noindent}if $(E_{n},\eta)$ has signature $(n-1,1)$ and

\begin{equation}
\| H\|^{2} \leq \hat{\Lambda}(n_{1},\ldots,n_{k})\ , \label{ineq2}
\end{equation}

{\par\noindent}if $(E_{n},\eta)$ has signature $(m-1,n-m+1)$.
\end{theorem}

{\par\noindent}\underline{Proof:}

{\par\noindent}Starting from the Gauss equation (\ref{gauss}) w.r.t. an orthonormal basis $\{e_{1},\ldots,e_{m},\xi_{m+1},\ldots,\xi_{n}\}$ we can express the scalar curvature of $M$ as

\begin{eqnarray}
2 R & = & \sum_{\alpha,\beta=1}^{m}\varepsilon_{\alpha\beta} R_{\alpha\beta\alpha\beta} \nonumber \\
    & = & \sum_{A=m+1}^{n}\signa\, \left(\sum_{\alpha=1}^{m}\varepsilon_{\alpha}\Omega_{\alpha\alpha}^{A}\right)^{2} - \sum_{A=m+1}^{n}\signa\, \sum_{\alpha,\beta=1}^{m}\varepsilon_{\alpha\beta}\left(\Omega_{\alpha\beta}^{A}\right)^{2} \nonumber \\
 & = & \| H\|^{2} - \Omega^{2}\ . \label{preq0}
\end{eqnarray}

{\par\noindent}If we put, with $k\geq 1$,

\begin{eqnarray*}
\phi & = & 2 R - \frac{m+k-1-\sum_{j=1}^{k}n_{j}}{m+k-\sum_{j=1}^{k}n_{j}} \| H\|^{2}\ , \\
\gamma & = & m+k-\sum_{j=1}^{k}n_{j}\ , 
\end{eqnarray*}

{\par\noindent}it is a small calculation to show that

\begin{equation}
\| H\|^{2} = \gamma (\phi +\Omega^{2}). \label{preq1}
\end{equation}

{\par\noindent}We choose $\xi_{m+1}$ along $\vec{H}$ and put $a_{\alpha}=\varepsilon_{\alpha}\Omega_{\alpha\alpha}^{m+1}$. Equation (\ref{preq1}) becomes

\begin{eqnarray}
 \lefteqn{\varepsilon_{m+1} \left(\sum_{\alpha=1}^{m}a_{\alpha}\right)^{2} = } \label{preq2} \\
  & &  \gamma \left\{ \phi +\varepsilon_{m+1}\sum_{\alpha=1}^{m}(a_{\alpha})^{2} +\varepsilon_{m+1}\sum_{\alpha\neq\beta=1}^{m}\varepsilon_{\alpha\beta} (\Omega_{\alpha\beta}^{m+1})^{2} +\sum_{A=m+2}^{n}\signa \sum_{\alpha,\beta=1}^{m}\varepsilon_{\alpha\beta} (\Omega_{\alpha\beta}^{A})^{2}\right\}\ . \nonumber
\end{eqnarray}

{\par\noindent}

{\par\noindent}If we use the notation

\begin{eqnarray*}
\bar{a}_{1} & = & a_{1}\ , \\
\bar{a}_{2} & = & a_{2}+...+a_{n_{1}}\ , \\
\bar{a}_{3} & = & a_{n_{1}+1}+...+a_{n_{1}+n_{2}}\ , \\
 \vdots & & \vdots \\
\bar{a}_{k+1} & = & a_{n_{1}+...+n_{k-1}+1}+...+a_{n_{1}+...+n_{k}}\ , \\
\bar{a}_{k+2} & = & a_{n_{1}+...+n_{k}+1}\ , \\
 \vdots & & \vdots \\
\bar{a}_{\gamma} & = & a_{m-1}\ , \\
\bar{a}_{\gamma+1} & = & a_{m}\ ,
\end{eqnarray*}

{\par\noindent}we have

\[ \left( \sum_{\alpha=1}^{\gamma+1}\bar{a}_{\alpha}\right)^{2} = \left( \sum_{\alpha=1}^{m}a_{\alpha}\right)^{2}\ , \]

{\par\noindent}and

\[  \sum_{\alpha=1}^{\gamma+1}(\bar{a}_{\alpha})^{2} = \sum_{\alpha=1}^{m}(a_{\alpha})^{2} + \sum_{2\leq \alpha_{1}\neq\beta_{1}\leq n_{1}} a_{\alpha_{1}}a_{\beta_{1}} + \sum_{\alpha_{2}\neq\beta_{2}\in Q_{2}} a_{\alpha_{2}}a_{\beta_{2}}+ ... + \sum_{\alpha_{k}\neq\beta_{k}\in Q_{k}} a_{\alpha_{k}}a_{\beta_{k}}\ , \]

{\par\noindent}with $Q_{1}=\{1,\ldots,n_{1}\}$, $Q_{2}=\{n_{1}+1,\ldots,n_{1}+n_{2}\}$, ..., $Q_{k}=\{n_{1}+\ldots+n_{k-1}+1,\ldots, n_{1}+\ldots+n_{k}\}$. Equation (\ref{preq2}) becomes

\begin{eqnarray}
\lefteqn{ \varepsilon_{m+1} \left(\sum_{\alpha=1}^{\gamma+1}\bar{a}_{\alpha}\right)^{2} = \gamma\left\{ \phi +\varepsilon_{m+1}\sum_{\alpha=1}^{\gamma+1}(\bar{a}_{\alpha})^{2} \right. } \nonumber \\
 & & +\varepsilon_{m+1}\sum_{\alpha\neq\beta=1}^{m}\varepsilon_{\alpha\beta}(\Omega_{\alpha\beta}^{m+1})^{2} +\sum_{A=m+2}^{n}\signa\sum_{\alpha,\beta=1}^{m}\varepsilon_{\alpha\beta}(\Omega_{\alpha\beta}^{A})^{2} \nonumber \\
 & & \left. -\varepsilon_{m+1}\sum_{2\leq\alpha_{1}\neq\beta_{1}\leq n_{1}}a_{\alpha_{1}}a_{\beta_{1}} - ... -\varepsilon_{m+1}\sum_{\alpha_{k}\neq\beta_{k}\in Q_{k}} a_{\alpha_{k}}a_{\beta_{k}} \right\} \label{preq3}
\end{eqnarray}

{\par\noindent}We need the following algebraic lemma,

\begin{lemma}[\onlinecite{chen2}]

{\par\noindent}If $\bar{a}_{1}, ..., \bar{a}_{n}, c$ are $n+1$ ($n\geq 2$) real numbers such that

\[ \left(\sum_{i=1}^{n}\bar{a}_{i}\right)^{2} = (n-1)\left( \sum_{i=1}^{n} (\bar{a}_{i})^{2} + c\right)\ , \]

{\par\noindent}we have that $2 \bar{a}_{1}\bar{a}_{2}\geq c$ and equality holds iff $\bar{a}_{1}+\bar{a}_{2}=\bar{a}_{3}=...=\bar{a}_{n}$.
\end{lemma}

{\par\noindent}Two seperate cases appear. We first look at the case when $\vec{H}$ is spacelike, i.e. $\varepsilon_{m+1}=1$. Using the above lemma equation (\ref{preq3}) becomes

\begin{eqnarray*}
\lefteqn{ \bar{a}_{1}\bar{a}_{2} \geq \frac{1}{2}\phi +\frac{1}{2}\sum_{\alpha\neq\beta=1}^{m}\varepsilon_{\alpha\beta}(\Omega_{\alpha\beta}^{m+1})^{2} +\frac{1}{2}\sum_{A=m+2}^{n}\signa \sum_{\alpha,\beta=1}^{m}\varepsilon_{\alpha\beta}(\Omega_{\alpha\beta}^{A})^{2} } \\
 & & -\frac{1}{2}\sum_{2\leq\alpha_{1}\neq\beta_{1}\leq n_{1}}a_{\alpha_{1}}a_{\beta_{1}} - ... -\frac{1}{2} \sum_{\alpha_{k}\neq\beta_{k}\in Q_{k}} a_{\alpha_{k}}a_{\beta_{k}}\ .
\end{eqnarray*}

{\par\noindent}Because

\[ \sum_{\alpha_{j}\neq\beta_{j}}a_{\alpha_{j}}a_{\beta_{j}} = 2 \sum_{\alpha_{j}<\beta_{j}}a_{\alpha_{j}}a_{\beta_{j}}\ , \]

{\par\noindent}we have

\begin{equation}
 \sum_{j=1}^{k}\sum_{\alpha_{j}<\beta_{j}\in Q_{j}}a_{\alpha_{j}}a_{\beta_{j}} \geq \frac{1}{2}\phi +\sum_{\alpha<\beta=1}^{m}\varepsilon_{\alpha\beta}(\Omega_{\alpha\beta}^{m+1})^{2} +\frac{1}{2}\sum_{A=m+2}^{n}\signa \sum_{\alpha,\beta=1}^{m}\varepsilon_{\alpha\beta}(\Omega_{\alpha\beta}^{A})^{2}\ . \label{preq4}
\end{equation}

{\par\noindent}Let $L_{j}$ be a $n_{j}$-dimensional subspace of $T_{p}M$ such that

\[ L_{j} =\, \mbox{span}\{e_{n_{1}+...+n_{j-1}+1}, ..., e_{n_{1}+...+n_{j}}\}\ . \]

{\par\noindent}The scalar curvature of the plane section is given by

\[ \tau(L_{j}) = \sum_{\alpha_{j}<\beta_{j}\in Q_{j}} \varepsilon_{\alpha_{j}\beta_{j}}\sum_{A=m+1}^{n}\signa \left[ \Omega_{\alpha_{j}\alpha_{j}}^{A}\Omega_{\beta_{j}\beta_{j}}^{A} - \left(\Omega_{\alpha_{j}\beta_{j}}^{A}\right)^{2} \right]\ . \]

{\par\noindent}Then using the above notation we find

\begin{eqnarray*}
 \lefteqn{ \tau(L_{1})+ ... +\tau(L_{k}) = \sum_{j=1}^{k}\sum_{\alpha_{j}<\beta_{j}\in Q_{j}} a_{\alpha_{j}}a_{\beta_{j}} } \\
  & & -\sum_{j=1}^{k}\sum_{\alpha_{j}<\beta_{j}\in Q_{j}}\varepsilon_{\alpha_{j}\beta_{j}} (\Omega_{\alpha_{j}\beta_{j}}^{m+1})^{2} +\sum_{j=1}^{k}\sum_{\alpha_{j}<\beta_{j}\in Q_{j}} \varepsilon_{\alpha_{j}\beta_{j}}\sum_{A=m+2}^{n}\signa \left[ \Omega_{\alpha_{j}\alpha_{j}}^{A}\Omega_{\beta_{j}\beta_{j}}^{A} - \left(\Omega_{\alpha_{j}\beta_{j}}^{A}\right)^{2}\right]\ .
\end{eqnarray*}

{\par\noindent}If we use the inequality (\ref{preq4}) and the notation

\begin{eqnarray*}
Q_{k+1}  & = & \{n_{1}+...+n_{k}+1, ..., m\}\ , \\
Q & = & Q_{1}\cup ...\cup Q_{k}\cup Q_{k+1}\ , \\
Q^{2} & = & (Q_{1}\times Q_{1})\cup ... \cup (Q_{k}\times Q_{k})\cup (Q_{k+1}\times Q_{k+1})\ , \\
\nabla^{2} & = & (Q\times Q)/ Q^{2}\ ,
\end{eqnarray*}

{\par\noindent}we have

\begin{eqnarray}
 \lefteqn{\tau(L_{1})+ ... +\tau(L_{k}) } \nonumber \\
 & & \geq\, \frac{1}{2}\phi +\frac{1}{2}\sum_{A=m+1}^{n}\signa \sum_{(\alpha,\beta)\in\nabla^{2}}\varepsilon_{\alpha\beta} (\Omega_{\alpha\beta}^{A})^{2} +\frac{1}{2}\sum_{A=m+2}^{n}\signa \sum_{j=1}^{k}\left( \sum_{\alpha\in Q_{j}}\varepsilon_{\alpha}\Omega_{\alpha\alpha}^{A}\right)^{2}\ . \label{preq5}
\end{eqnarray}

{\par\noindent}The signature of the embedding space $E_{n}$ is chosen to be $(n-1,1)$ such that all $\signa=1$ and the condition of causal-type preserving ensures that the terms with possible minus signs appearing on the righthand side vanish. We have

\[ \tau(L_{1})+...+\tau(L_{k}) \geq \frac{1}{2}\phi\ . \]

{\par\noindent}This holds for all mutually orthogonal subspaces $L_{j}$, in particular for the infimum,

\begin{equation}
\| H\|^{2} \geq \Lambda(n_{1},...,n_{k})\ .
\end{equation}

{\par\noindent}The case when $\vec{H}$ is timelike is analogous and we find instead of (\ref{preq5}),

\begin{eqnarray*}
 \lefteqn{ \tau(L_{1})+...+\tau(L_{k}) } \\
 & & \leq \frac{1}{2}\phi +\frac{1}{2}\sum_{A=m+1}^{n}\varepsilon_{A}\sum_{(\alpha,\beta)\in\nabla^{2}} \varepsilon_{\alpha\beta}(\Omega_{\alpha\beta}^{A})^{2} +\frac{1}{2}\sum_{A=m+2}^{n}\varepsilon_{A}\sum_{j=1}^{k} \left( \sum_{\alpha_{j}\in Q_{j}}\varepsilon_{\alpha_{j}}\Omega_{\alpha_{j}\alpha_{j}}^{A}\right)^{2}\ .
\end{eqnarray*}

{\par\noindent}We choose the signature of the embedding space to be $(m-1,n-m+1)$, i.e. all normal directions are timelike. We find

\[ \tau(L_{1})+ ... +\tau(L_{k}) \leq \frac{1}{2}\phi\ . \]

{\par\noindent}This holds again for all mutually orthogonal subspaces, in particular for the supremum,

\begin{equation}
\| H\|^{2} \leq \hat{\Lambda}(n_{1},...,n_{k})\ .
\end{equation}

{\par\noindent}It remains to show the inequality when $k=0$. Starting from (\ref{preq0}) and again choosing $\xi_{m+1}$ along $\vec{H}$ we find

\begin{equation}
 2R = \| H\|^{2} -\varepsilon_{m+1}\sum_{\alpha=1}^{m}(a_{\alpha})^{2} -\varepsilon_{m+1} \sum_{\alpha\neq\beta=1}^{m}\varepsilon_{\alpha\beta}(\Omega_{\alpha\beta}^{m+1})^{2} -\sum_{A=m+2}^{n}\signa \sum_{\alpha,\beta=1}^{m}\varepsilon_{\alpha\beta}(\Omega_{\alpha\beta}^{A})^{2}\ , \label{preq6}
\end{equation}

{\par\noindent}with $a_{\alpha}=\varepsilon_{\alpha}\Omega_{\alpha\alpha}^{m+1}$. We have

\begin{eqnarray*}
\sum_{\alpha=1}^{m}(a_{\alpha})^{2} & = & \left( \sum_{\alpha=1}^{m}a_{\alpha}\right)^{2} -2\sum_{\alpha<\beta=1}^{m} a_{\alpha}a_{\beta} \\
 & = & \varepsilon_{m+1}\| H\|^{2} +\sum_{\alpha<\beta=1}^{m}(a_{\alpha}-a_{\beta})^{2} - (m-1)\sum_{\alpha=1}^{m}(a_{\alpha})^{2} \\
m\sum_{\alpha=1}^{m}(a_{\alpha})^{2} & = & \varepsilon_{m+1}\| H\|^{2} +\sum_{\alpha<\beta=1}^{m}(a_{\alpha}-a_{\beta})^{2} \\
 & \geq & \varepsilon_{m+1}\| H\|^{2}\ .
\end{eqnarray*}

{\par\noindent}If $\vec{H}$ is spacelike, (\ref{preq6}) with the above inequality becomes

\[  2R \leq \frac{m-1}{m} \| H\|^{2} -\sum_{\alpha\neq\beta=1}^{m}\varepsilon_{\alpha\beta} (\Omega_{\alpha\beta}^{m+1})^{2} -\sum_{A=m+2}^{n}\signa \sum_{\alpha,\beta=1}^{m}\varepsilon_{\alpha\beta} (\Omega_{\alpha\beta}^{A})^{2}\ . \]

{\par\noindent}The signature of the embedded space is chosen to be $(n-1,1)$ and because of the condition of causal-type preserving, we find

\begin{equation}
\| H\|^{2} \geq \frac{2m}{m-1}R = \Lambda(0)\ .
\end{equation}

{\par\noindent}The proof for the timelike case is similar.\hfill $\diamond$

\vspace{4mm}

{\par\noindent}Notice that due to our choice of signature for the embedded space $E_{n}$, $\vec{H}$ is always non-null. So we exclude the case of quasi-minimal embeddings.

{\par\noindent}If there is equality we can determine the form of the second fundamental forms.

\begin{corollary}
There is equality in (\ref{ineq1}) or (\ref{ineq2}) at a point $p \in M$ iff there exists an orthonormal basis at $p$ such that the second fundamental forms take the form

\[ \Omega_{m+1} = \left( \begin{array}{cccc}
                    a_{1} &       &        &       \\
                          & a_{2} &        &       \\
                          &       & \ddots &       \\
                          &       &        & a_{n} \\
                    \end{array} \right)\ , \]

{\par\noindent}with $a_{1}+...+a_{n_{1}}= a_{n_{1}+1}+...+a_{n_{1}+n_{2}}=...=a_{n_{1}+...+n_{k-1}+1}+...+a_{n_{1}+...+n_{k}}= a_{n_{1}+...+n_{k}+1}=...=a_{m}$ and

\[ \Omega_{r} = \left( \begin{array}{ccccccc}
                  A_{r 1} &        &        &         &   &        &  \\
                         & A_{r 2} &        &         &   &        &  \\
                         &        & \ddots &         &   &        &  \\
                         &        &        & A_{r k} &   &        &  \\
                         &        &        &         & 0 &        &  \\  
                         &        &        &         &   & \ddots &  \\
                         &        &        &         &   &        & 0 \\
                 \end{array} \right)\ , \]

{\par\noindent}with $Trace(A_{r j}) = 0$, $r=m+2,...,n$, $j=1,...,k$.
\end{corollary} 

{\par\noindent}As in Ref.~\onlinecite{chen} we have the following

\begin{definition}
An isometric embedding $x: (M,g)\rightarrow (E_{n},\eta)$ is called an ideal embedding if and only if there exists a k-tuple $(n_{1},\ldots,n_{k})$ such that in a neighbourhood $U$ of a point $p\in M$ there is equality in (\ref{ineq1}) or (\ref{ineq2}) respectively.
\end{definition}

{\par\noindent}If the pseudo-Euclidean embedding space has signature $(n-1,1)$ an ideally embedded manifold $M$ means that the squared mean curvature of $M$ is minimal. Because $\vec{H}$ measures the tension on $M$ from the surrounding space an ideal embedding in $E_{n-1,1}$ can be considered as {\it a best way of living in a best world} for the neighbourhood $U$ \cite{chen}. In the case of an embedding space with signature $(n-m+1,m-1)$ the situation is reversed. An ideally embedded manifold receives the maximum possible amount of tension from the surrounding space at each point of $M$. Although this situation is not {\it ideal} we reserve the notation for both occasions.


\section{Ideally embedded space-times}

{\par\noindent}Using the above notion of ideal embedding gives us a natural set of second fundamental forms to consider. Notice that this is the reverse situation usually adopted in the literature. There one often starts from a given metric and looks for the minimal embedding, i.e. with the least extra dimensions, or one puts some constraints on the curvature tensor through the choice of matter and/or Petrov type (Although see Ref.~\onlinecite{vandenb} for a different approach).

{\par\noindent}We will restrict our manifold $M$ to be a four-dimensional space-time embedded in a five-dimensional pseudo-Euclidean space. The torsion vector is zero in this case, so (\ref{ricci}) is trivially satisfied and (\ref{codazzi}) simplifies significantly. We further only study those cases when there is equality for a k-tuple with only  spacelike plane sections. The case with a timelike plane section in the k-tuple will be considered separately.

{\par\noindent}We denote the orthonormal basis of an ideally embedded space-time $M$ for which the second fundamental forms take their special forms as $\{e_{\alpha}\} = \{\vec{w},\vec{v},\vec{q},\vec{u}\}$ with $u_{\alpha}u^{\alpha}=-1$. From the above corollary we have three possible cases:

{\par\noindent}i) equality with $k=0$,

\[ \Omega_{\alpha\beta}= \mu g_{\alpha\beta}\ , \]

{\par\noindent}ii) equality with $k=1$, $n=2$,

\[ \Omega_{\alpha\beta} = (\mu-\lambda)w_{\alpha}w_{\beta}+\lambda v_{\alpha}v_{\beta} +\mu q_{\alpha}q_{\beta} -\mu u_{\alpha}u_{\beta}\ , \]

{\par\noindent}iii) equality with $k=1$, $n=3$,

\[ \Omega_{\alpha\beta} = (\mu-\lambda-\nu)w_{\alpha}w_{\beta} +\lambda v_{\alpha}v_{\beta} +\nu q_{\alpha}q_{\beta} -\mu u_{\alpha}u_{\beta}\ . \]

{\par\noindent}Before we determine the metrics which can be ideally embedded with one of the above second fundamental forms we mention two results which limit the possible outcomes.

\begin{theorem}[\onlinecite{szekeres}] \label{thszek}
No nonflat vacuum metric can be embedded in a 5-dimensional pseudo-Euclidean space. 
\end{theorem}

\begin{theorem}[\onlinecite{collinson}] \label{thcollin}
There are no embedding class one solutions of the Einstein-Maxwell equations with a non-null electromagnetic field.
\end{theorem}


\subsection{Case i:}

{\par\noindent}If we take as shape operator

\[ \Omega_{\alpha\beta} = \mu g_{\alpha\beta}\ , \]

{\par\noindent}i.e. the embedding is umbilical, the Codazzi equations (\ref{codazzi}) become

\[ g_{\alpha\beta}\nabla_{\gamma}\mu = g_{\alpha\gamma}\nabla_{\beta}\mu\ , \]

{\par\noindent}or contracting over $\alpha$ and $\beta$ gives $\nabla_{\gamma}\mu=0$. The Gauss equations (\ref{gauss}) give

\[ R_{\alpha\beta\gamma\delta} = 2\varepsilon\mu^{2} g_{\alpha[\gamma}g_{\delta]\beta}\ , \]

{\par\noindent}with $\mu$ a constant. The space-time is a space of constant curvature, a de Sitter space if $\varepsilon=1$ or an anti de Sitter space if $\varepsilon=-1$ (Ref.~\onlinecite{stephani} p103). Due to our assumption of time-orientability the space obtained from the de Sitter space in which points are identified by reflection through the origin of the embedding space is excluded (Ref.~\onlinecite{hawking} p130).


\subsection{Case ii:}

{\par\noindent}With respect to the orthonomal basis $\{w^{\alpha}, v^{\alpha}, q^{\alpha}, u^{\alpha}\}$, $u^{\alpha}$ timelike, the second fundamental form becomes,

\[ \Omega_{\alpha\beta} = -\mu u_{\alpha}u_{\beta}+\mu q_{\alpha}q_{\beta} +\lambda v_{\alpha}v_{\beta}+(\mu-\lambda)w_{\alpha}w_{\beta}\ . \]

{\par\noindent}If we decompose the covariant derivatives,

\begin{eqnarray*}
\nabla_{\beta}u_{\alpha} & = & w_{\alpha}A_{\beta} +v_{\alpha}B_{\beta} +q_{\alpha}C_{\beta}\ , \\
\nabla_{\beta}w_{\alpha} & = & u_{\alpha}A_{\beta} +v_{\alpha}D_{\beta} +q_{\alpha}E_{\beta}\ , \\
\nabla_{\beta}v_{\alpha} & = & u_{\alpha}B_{\beta} - w_{\alpha}D_{\beta} +q_{\alpha}F_{\beta}\ , \\
\nabla_{\beta}q_{\alpha} & = & u_{\alpha}C_{\beta} - w_{\alpha}E_{\beta} - v_{\alpha}F_{\beta}\ ,
\end{eqnarray*}

{\par\noindent}the Codazzi equations give

\begin{eqnarray*}
\lambda A_{\alpha} & = & \nabla_{w}\mu\, u_{\alpha} +\lambda A_{v}\, v_{\alpha} -\nabla_{u}\lambda\, w_{\alpha}\ , \\
(\lambda-\mu)B_{\alpha} & = & -\nabla_{v}\mu\, u_{\alpha} - \nabla_{u}\lambda\, v_{\alpha} -\lambda A_{v}\, w_{\alpha}\ , \\
(2\lambda-\mu)D_{\alpha} & = & \lambda A_{v}\, u_{\alpha}+ (2\lambda-\mu)D_{q}\, q_{\alpha} -\nabla_{w}\lambda\, v_{\alpha} -\nabla_{v}(\mu-\lambda)\, w_{\alpha}\ , \\
\lambda E_{\alpha} & = & -\nabla_{w}\mu\, q_{\alpha}+ (2\lambda-\mu)D_{q}\, v_{\alpha} +\nabla_{q}\lambda\, w_{\alpha}\ , \\
(\lambda-\mu)F_{\alpha} & = & \nabla_{v}\mu\, q_{\alpha}+\nabla_{q}\lambda\, v_{\alpha} -(2\lambda-\mu)D_{q}\, w_{\alpha}\ , 
\end{eqnarray*}

{\par\noindent}and

\begin{equation}
\nabla_{u}\mu\, =\, \nabla_{q}\mu\, =\, 0\ , 
\end{equation}

{\par\noindent}with $A_{v}, D_{q}$ scalars and $u^{\alpha}\nabla_{\alpha}=\nabla_{u}$, etc. There is no equation for $C_{\alpha}$.

{\par\noindent}The Ricci identities $2\nabla_{[\gamma}\nabla_{\beta]}z_{\alpha}= z^{\sigma}R_{\sigma\alpha\beta\gamma}$, with $z^{\alpha}$ one of the basis vectors, give

\begin{eqnarray}
\nabla_{[\alpha}A_{\beta]} -D_{[\alpha}B_{\beta]} -E_{[\alpha}C_{\beta]} & = & \varepsilon\mu(\mu-\lambda)\, u_{[\beta}w_{\alpha]}\ , \label{RI21} \\
\nabla_{[\alpha}B_{\beta]}+D_{[\alpha}A_{\beta]}-F_{[\alpha}C_{\beta]} & = & \varepsilon\mu\lambda\,  u_{[\beta}v_{\alpha]}\ , \\
\nabla_{[\alpha}C_{\beta]}+E_{[\alpha}A_{\beta]}+F_{[\alpha}B_{\beta]} & = & \varepsilon\mu^{2}\, u_{[\beta}q_{\alpha]}\ , \\
\nabla_{[\alpha}D_{\beta]}+B_{[\alpha}A_{\beta]}-F_{[\alpha}E_{\beta]} & = & \varepsilon\lambda(\mu-\lambda)\, w_{[\beta}v_{\alpha]}\ , \\
\nabla_{[\alpha}E_{\beta]}+C_{[\alpha}A_{\beta]}+F_{[\alpha}D_{\beta]} & = & \varepsilon\mu(\mu-\lambda)\, w_{[\beta}q_{\alpha]}\ , \\
\nabla_{[\alpha}F_{\beta]}+C_{[\alpha}B_{\beta]}-E_{[\alpha}D_{\beta]} & = & \varepsilon\mu\lambda\,  v_{[\beta}q_{\alpha]}\ .
\end{eqnarray}


\subsubsection{If $\lambda=\mu\neq 0$:}

{\par\noindent}From the Codazzi equations we find $\nabla_{v}\mu\, =\, A_{v}\, =\, D_{q}\, =\, 0$ and 

\[ \nabla_{\beta}w_{\alpha} = \nabla_{w}\ln\lambda\, (u_{\alpha}u_{\beta}-v_{\alpha}v_{\beta}-q_{\alpha}q_{\beta})\ . \]

{\par\noindent}Let us denote the projection operator on the timelike hypersurface orthogonal to $w^{\alpha}$ by $h_{\alpha}^{\ \beta} = \delta_{\alpha}^{\ \beta} - w_{\alpha}w^{\beta}$. From the Gauss equations we find that

\[ w^{\alpha}R_{\alpha\beta\gamma\delta} = 0\ , \]

{\par\noindent}and so $w^{\alpha}$ is a constant vector field (see Ref.~\onlinecite{stephani} p553), i.e. $\nabla_{\beta}w_{\alpha}=0$ or $\lambda=\mu$ =constant. If we denote with $^{3}\!R_{\alpha\beta\gamma\delta}$ the Riemann tensor of the timelike hypersurface, the Gauss equations give

\[ ^{3}\!R_{\alpha\beta\gamma\delta}= R_{\alpha\beta\gamma\delta} = 2\varepsilon\lambda^{2} h_{\alpha[\gamma}h_{\delta]\beta}\ . \]

{\par\noindent}The timelike 3-space is a space of constant curvature. We can then choose coordinates such that the metric reads

\begin{equation}
\ds^{2} = \mbox{d}z^{2} +\frac{\mbox{d}y^{2}+\mbox{d}x^{2}-\mbox{d}t^{2}}{[ 1+\frac{1}{4}\varepsilon\lambda^{2}(y^{2}+x^{2}-t^{2})]^{2}}\ , 
\end{equation}

{\par\noindent}with $\lambda$=constant. Because the embedding is quasi-umbilical (i.e. there exist functions $\phi$ and $\psi$ such that $\Omega_{\alpha\beta}=\phi g_{\alpha\beta}+\psi w_{\alpha}w_{\beta}$) the metric is conformally flat \cite{deszcz}. The Ricci tensor is 

\[ R_{\alpha\beta} = 2\varepsilon\lambda^{2} h_{\alpha\beta}\ , \]

{\par\noindent}with Segr\'{e} type  A1, $[1 (1 1, 1)]$, and the energy-momentum tensor does not satisfies any of the known energy conditions \cite{hawking}. Due to the observation that the Universe is accelerating, cosmological models with such a strange equation of state are recently under investigation.


\subsubsection{If $\lambda=0$, $\mu\neq 0$:}

{\par\noindent}From the Codazzi equations we find $\nabla_{w}\mu =D_{q} = 0$ and 

\[ \nabla_{\beta}v_{\alpha} = \nabla_{v}\ln\mu\, (u_{\alpha}u_{\beta}-w_{\alpha}w_{\beta}-q_{\alpha}q_{\beta})\ . \]

{\par\noindent}This is the previous case with the roles of $v^{\alpha}$ and $w^{\alpha}$ interchanged.

\subsubsection{If $\mu=2\lambda\neq 0$:}

{\par\noindent}The Codazzi equations give $A_{v}=\nabla_{v}\lambda=\nabla_{w}\lambda = 0$. Then $A_{\alpha}=B_{\alpha}=D_{\alpha}=E_{\alpha}=F_{\alpha}=0$. The Ricci identity (\ref{RI21}) gives $\lambda=0$, so we must take $\mu\neq 2\lambda$.


\subsubsection{If $\lambda\neq 0$, $\mu-\lambda\neq 0$ and $\mu-2\lambda\neq 0$:}

{\par\noindent}Let $p_{\alpha}^{\ \beta}= \delta_{\alpha}^{\ \beta}- v_{\alpha}v^{\beta} -w_{\alpha}w^{\beta}$ be the projection operator on the 2-space $V_{2}$ orthogonal to $v^{\alpha}$ and $w^{\alpha}$. The second fundamental forms of the embedding of $V_{2}$ in the space-time $(M,g)$ are

\[ \Omega^{v}_{\alpha\beta} = p_{(\alpha}^{\ \gamma}p_{\beta)}^{\ \sigma}\nabla_{\gamma}v_{\sigma} = \frac{\nabla_{v}\mu}{\lambda-\mu}p_{\alpha\beta}\ , \]

{\par\noindent}and 

\[ \Omega^{w}_{\alpha\beta} = -\frac{\nabla_{w}\mu}{\lambda}p_{\alpha\beta}\ . \]

{\par\noindent}Using the Gauss equations we find for the Riemann tensor of the timelike 2-space $V_{2}$,

\begin{equation}
^{2}\!R_{\alpha\beta\gamma\delta} = 2\left\{ \varepsilon\mu^{2} +\left(\frac{\nabla_{v}\mu}{\lambda-\mu}\right)^{2} +\left(\frac{\nabla_{w}\mu}{\lambda}\right)^{2}\right\} p_{\alpha[\gamma}p_{\delta]\beta}\ .
\end{equation}

{\par\noindent}It is a small calculation to show that the coefficient has zero-derivative in the $u$ and $q$ directions. The 2-space $V_{2}$ is a space of constant curvature. We can choose coordinates such that

\[ w_{\alpha} = (e^{\phi(y,z)},0,0,0)\ \mbox{  ,  }\ v_{\alpha}=(0,e^{\xi(y,z)},0,0)\ , \]

{\par\noindent}and the metric reads

\begin{equation}
\ds^{2} = e^{2\phi(y,z)}\mbox{d}z^{2} +e^{2\xi(y,z)}\mbox{d}y^{2} +Y^{2}(y,z) \{ \mbox{d}x^{2} -\Sigma^{2}(x,k)\mbox{d}t^{2}\}\ ,
\end{equation}

{\par\noindent}with $\Sigma(x,k)= \sin(x), x\ \mbox{or}\ \sinh(x)$ if $k=1,0\ \mbox{or}\ -1$ and 

\[ kY^{-2} = \varepsilon\mu^{2} +\left(\frac{\nabla_{v}\mu}{\lambda-\mu}\right)^{2} +\left( \frac{\nabla_{w}\mu}{\lambda}\right)^{2}\ . \]

{\par\noindent}These metrics have a group $G_{3}$ working on the two-surface of constant curvature and therefore have Petrov type D or O. Because the two-surface is timelike the energy-momentum content cannot be a perfect fluid, a null electromagnetic field or pure radiation (see Ref.~\onlinecite{stephani} ch.15) and due to theorems \ref{thszek} and \ref{thcollin} also vacuum and an electromagnetic non-null field are not possible. We can however interpret this space-time as filled with an anisotropic perfect fluid satisfying the strong energy condition if and only if the extra dimension is timelike ($\varepsilon=-1$) and $\mu$ and $\lambda$ satisfy any of the following conditions:

\[ 1)\ \lambda > 0\ ,\ \mu > \lambda\ , \]
\[ 2)\ \lambda > 0\, ,\ -\lambda\leq\mu\leq \frac{1}{2}\lambda\ , \]
\[ 3)\ \lambda < 0\, ,\ \frac{1}{2}\lambda\leq\mu\leq -\lambda\ , \]
\[ 4)\ \lambda < 0\, ,\ \mu < \lambda\ . \]


\subsection{Case iii:}

{\par\noindent}With respect to an orthonormal tetrad $\{w^{\alpha}, v^{\alpha}, q^{\alpha}, u^{\alpha}\}$ the shape operator takes the form

\begin{equation}
\Omega_{\alpha\beta} = -\mu u_{\alpha}u_{\beta}+ \nu q_{\alpha}q_{\beta} +\lambda v_{\alpha}v_{\beta} +(\mu-\lambda-\nu)w_{\alpha}w_{\beta}\ .
\end{equation}

{\par\noindent}If we use the same decompositions of the covariant derivatives as in the previous case, the Codazzi equations give

\begin{eqnarray*}
 (\lambda+\nu)A_{\alpha} & = & \nabla_{w}\mu\, u_{\alpha} +\nabla_{u}(\mu-\lambda-\nu)\, w_{\alpha} +(\lambda+\nu)A_{v}\, v_{\alpha} +(\lambda+\nu)A_{q}\, q_{\alpha}\ , \\
 (\mu-\lambda)B_{\alpha} & = & -\nabla_{v}\mu\, u_{\alpha}+ (\lambda+\nu)A_{v}\, w_{\alpha}+\nabla_{u}\lambda\, v_{\alpha} +(\mu-\lambda)B_{q}\, q_{\alpha}\ , \\
 (\mu-\nu)C_{\alpha} & = & -\nabla_{q}\mu\, u_{\alpha} +(\lambda+\nu)A_{q}\, w_{\alpha} +(\mu-\lambda)B_{q}\, v_{\alpha} +\nabla_{u}\nu\, q_{\alpha}\ , \\
 (\mu-2\lambda-\nu)D_{\alpha} & = & -(\lambda+\nu)A_{v}\, u_{\alpha}+\nabla_{v}(\mu-\lambda-\nu)\, w_{\alpha} +\nabla_{w}\lambda\, v_{\alpha} +(\mu-2\lambda-\nu)D_{q}\, q_{\alpha}\ , \\
 (\mu-\lambda-2\nu)E_{\alpha} & = & -(\lambda+\nu)A_{q}\, u_{\alpha} +\nabla_{q}(\mu-\lambda-\nu)\, w_{\alpha}+(\mu-2\lambda-\nu)D_{q}\, v_{\alpha} +\nabla_{w}\nu\, q_{\alpha}\ , \\
 (\lambda-\nu)F_{\alpha} & = & -(\mu-\lambda)B_{q}\, u_{\alpha} +(\mu-2\lambda-\nu)D_{q}\, w_{\alpha} +\nabla_{q}\lambda\, v_{\alpha} +\nabla_{v}\nu\, q_{\alpha}\ ,
\end{eqnarray*}

{\par\noindent}with $A_{v}, A_{q}, B_{q}, D_{q}$ scalars. The Ricci identities are

\begin{eqnarray}
\nabla_{[\alpha}A_{\beta]} -D_{[\alpha}B_{\beta]} -E_{[\alpha}C_{\beta]} & = & \varepsilon\mu(\mu-\lambda-\nu)\, u_{[\beta}w_{\alpha]}\ , \label{RI31} \\
\nabla_{[\alpha}B_{\beta]}+D_{[\alpha}A_{\beta]}-F_{[\alpha}C_{\beta]} & = & \varepsilon\mu\lambda\, u_{[\beta}v_{\alpha]}\ , \label{RI32} \\
\nabla_{[\alpha}C_{\beta]}+E_{[\alpha}A_{\beta]}+F_{[\alpha}B_{\beta]} & = & \varepsilon\mu\nu\,  u_{[\beta}q_{\alpha]}\ , \label{RI33} \\
\nabla_{[\alpha}D_{\beta]}+B_{[\alpha}A_{\beta]}-F_{[\alpha}E_{\beta]} & = & \varepsilon\lambda(\mu-\lambda-\nu)\, w_{[\beta}v_{\alpha]}\ , \label{RI34} \\
\nabla_{[\alpha}E_{\beta]}+C_{[\alpha}A_{\beta]}+F_{[\alpha}D_{\beta]} & = & \varepsilon\nu(\mu-\lambda-\nu)\,  w_{[\beta}q_{\alpha]}\ , \label{RI35} \\
\nabla_{[\alpha}F_{\beta]}+C_{[\alpha}B_{\beta]}-E_{[\alpha}D_{\beta]} & = & \varepsilon\nu\lambda\,  v_{[\beta}q_{\alpha]}\ . \label{RI36}
\end{eqnarray}

{\par\noindent}Using the Gauss equations we find the Ricci tensor,

\begin{equation}
 R_{\alpha\beta} = \varepsilon\{ -\mu^{2}u_{\alpha}u_{\beta} +\nu(2\mu-\nu)q_{\alpha}q_{\beta}+ \lambda(2\mu-\lambda)v_{\alpha}v_{\beta}+ (\mu-\lambda-\nu)(\mu+\lambda+\nu)w_{\alpha}w_{\beta}\}\ . \label{ricciten}
\end{equation}

{\par\noindent}In the generic case the Segr\'{e} type is A1, $[1 1 1, 1]$. We will restrict the calculations in the following to perfect fluid space-times. This means $\mu,\lambda$ and $\nu$ must satisfy one of the following conditions

\begin{eqnarray*}
A)\ \ \mu=3\lambda & , & \nu=\lambda\ , \\
B)\ \ \mu=-\lambda & , & \nu=\lambda\ , \\
C)\ \ \mu=-\nu & , & \lambda = -3\nu\ , \\
D)\ \ \mu=-\lambda & , & \nu=-3\lambda\ .
\end{eqnarray*}

{\par\noindent}The cases $B, C$ and $D$ are the same with the roles of the spacelike vectors interchanged. Before we study the above cases in detail we give first the decomposition of the covariant derivative of the timelike direction $u^{\alpha}$ into its irreducible parts if $\lambda+\nu\neq 0$, $\mu-\lambda\neq 0$ and $\mu-\nu\neq 0$.

{\par\noindent}The acceleration reads,

\begin{equation}
\dot{u}_{\alpha} = \frac{\nabla_{q}\mu}{\mu-\nu}q_{\alpha}+\frac{\nabla_{v}\mu}{\mu-\lambda}v_{\alpha} -\frac{\nabla_{w}\mu}{\lambda+\nu}w_{\alpha}\ , 
\end{equation}

{\par\noindent}the expansion,

\begin{equation}
\theta = \frac{\nabla_{u}(\mu-\lambda-\nu)}{\lambda+\nu} +\frac{\nabla_{u}\lambda}{\mu-\lambda} +\frac{\nabla_{u}\nu}{\mu-\nu}\ , 
\end{equation}

{\par\noindent}the shear,

\begin{eqnarray*}
 \lefteqn{\sigma_{\alpha\beta} = \left\{\frac{2\nabla_{u}(\mu-\lambda-\nu)}{3(\lambda+\nu)} -\frac{\nabla_{u}\lambda}{3(\mu-\lambda)}-\frac{\nabla_{u}\nu}{3(\mu-\nu)}\right\}w_{\alpha}w_{\beta} +\frac{(\mu+\nu)A_{v}}{\mu-\lambda}w_{(\alpha}v_{\beta)} } \\
 & & +\frac{(\mu+\lambda)A_{q}}{\mu-\nu}w_{(\alpha}q_{\beta)} + \left\{ -\frac{\nabla_{u}(\mu-\lambda-\nu)}{3(\lambda+\nu)} +\frac{2\nabla_{u}\lambda}{3(\mu-\lambda)}-\frac{\nabla_{u}\nu}{3(\mu-\nu)}\right\} v_{\alpha}v_{\beta} \\
 & & +\frac{(2\mu-\lambda-\nu)B_{q}}{\mu-\nu}v_{(\alpha}q_{\beta)} +\left\{ -\frac{\nabla_{u}(\mu-\lambda-\nu)}{3(\lambda+\nu)} -\frac{\nabla_{u}\lambda}{3(\mu-\lambda)}+\frac{2\nabla_{u}\nu}{3(\mu-\nu)}\right\} q_{\alpha}q_{\beta}\ , 
\end{eqnarray*}

{\par\noindent}and the vorticity,

\begin{equation}
 \omega_{\alpha\beta} = \frac{(\mu-2\lambda-\nu)A_{v}}{\mu-\lambda}w_{[\alpha}v_{\beta]} +\frac{(\mu-\lambda-2\nu)A_{q}}{\mu-\nu}w_{[\alpha}q_{\beta]} +\frac{(\lambda-\nu)B_{q}}{\mu-\nu}v_{[\alpha}q_{\beta]}\ .
\end{equation}


\subsubsection{If $\mu=-\lambda$ and $\nu=\lambda\neq 0$:}

{\par\noindent}From the Codazzi equations we find $B_{q}=D_{q}=0$ and $\nabla_{v}\lambda=\nabla_{q}\lambda=0$. Projecting the Ricci identity (\ref{RI32}) on $u^{\alpha}v^{\beta}$ and (\ref{RI33}) on $u^{\alpha}q^{\beta}$ gives $A_{v}^{2} = A_{q}^{2}$. If we further project (\ref{RI32}) on $u^{\alpha}q^{\beta}$ we find $A_{v}A_{q}=0$, so

\[ A_{v} = A_{q} = 0\ . \]

{\par\noindent}Combining (\ref{RI31}), (\ref{RI32}) and (\ref{RI34}) gives

\begin{equation} 
(\nabla_{w}\ln\lambda)^{2} = 4(\nabla_{u}\ln\lambda)^{2}\ , \label{defvgl2}
\end{equation}

\begin{equation}
\nabla_{u}\nabla_{w}\ln\lambda = \frac{3}{2}\nabla_{u}\ln\lambda\, \nabla_{w}\ln\lambda\ , \label{defvgl4}
\end{equation}

{\par\noindent}and (\ref{RI32}) then becomes

\begin{equation}
2\nabla_{u}\nabla_{u}\ln\lambda -3 (\nabla_{u}\ln\lambda)^{2} -4\varepsilon\lambda^{2} = 0\ . \label{defvgl3}
\end{equation}

{\par\noindent}If we differentiate (\ref{defvgl2}) in the direction of $u^{\alpha}$ and use (\ref{defvgl4}) and (\ref{defvgl3}) we find $\lambda=0$. This case does not lead to ideally embedded perfect fluid space-times.


\subsubsection{If $\mu=3\lambda$ and $\nu=\lambda$:}

{\par\noindent}From the Codazzi equations we find $A_{v}=A_{q}=B_{q}=0$ and $\nabla_{w}\lambda=\nabla_{v}\lambda=\nabla_{q}\lambda=0$. We find that $u^{\alpha}$ is geodesic, hypersurface orthogonal and shearfree. The expansion of the timelike congruence with tangent $u^{\alpha}$ is given by $\theta=\frac{3}{2}\nabla_{u}\ln\lambda$. From the Ricci identities we have the equation

\begin{equation}
2\nabla_{u}\nabla_{u}\ln\lambda +(\nabla_{u}\ln\lambda)^{2} -12\varepsilon\lambda^{2} = 0\ . \label{defvgl}
\end{equation}

{\par\noindent}It follows that if $\theta=0$, $\lambda=0$ and space-time is flat. Therefore we take $\theta\neq 0$. We then choose coordinates adapted to the timelike vector, $u_{\alpha}=(0,0,0,u_{4})$. The metric becomes

\[ \ds^{2} = h_{ij}\mbox{d}x^{i}\mbox{d}x^{j} - (u_{4})^{2}\mbox{d}t^{2}\ , \]

{\par\noindent}with $i,j=1,2,3$. Then $\theta=\theta(t)$, $\lambda=\lambda(t)$ and $u_{4}=u_{4}(t)$. The second fundamental form of the embedding of the spacelike hypersurface orthogonal to $u^{\alpha}$ in $(M,g)$ is

\[ \Omega^{u}_{ij} = \frac{1}{3}\theta h_{ij}\ . \]

{\par\noindent}The Riemann tensor of the 3-space reads

\[ ^{3}\!R_{ijkl} = 2\left\{\varepsilon\lambda^{2} -\frac{1}{9}\theta^{2}\right\}h_{i[k}h_{l]j}\ , \]

{\par\noindent}the spacelike hypersurface is a space of constant curvature. The metric can be written, after a coordinate transformation $u_{4}(t)\dt\rightarrow \dt$, as 

\begin{equation}
\ds^{2} = a^{2}(t)\{ \mbox{d}r^{2}+\Sigma^{2}(r,k) (\mbox{d}\phi^{2} +\sin^{2}(\phi) \mbox{d}\psi^{2} )\} -\mbox{d}t^{2}\ , 
\end{equation}

{\par\noindent}with 

\begin{equation}
k a^{-2}=\varepsilon\lambda^{2}-\frac{1}{9}\theta^{2}\ , \label{defa}
\end{equation}

{\par\noindent}and $\Sigma(r,k)=\sin(r), r\ \mbox{or}\ \sinh(r)$ if $k=1, 0\, \mbox{or}\, -1$. This metric is a Robertson-Walker metric (see Ref.~\onlinecite{robertson} for the first results on the embedding of R-W models in flat 5-dimensional spaces). Combining (\ref{defvgl}) and (\ref{defa}) $a(t)$ must be a solution of

\begin{equation}
(\partial_{t}a)^{2} = \varepsilon c^{2}a^{6} - k\ ,
\end{equation}

{\par\noindent}with $c$ =constant and $\lambda = c a^{2}$. From the expression of the Ricci tensor (\ref{ricciten}) and the Einstein equations we can write the energy $\rho$ and pressure $p$ of the perfect fluid as

\[ \kappa\rho = 3 c^{2} a^{4}\ \mbox{and}\ \kappa p = -7 c^{2} a^{4}\ , \]

{\par\noindent}with $\varepsilon=+1$.


\section{Conclusion}

{\par\noindent}In the study of embeddings of a space-time in some higher-dimensional space attention has focused primarily on intrinsic properties of the submanifold (e.g. the source type or Petrov type). But the fact that we embed our space-time metric in a greater space gives us the opportunity to consider also extrinsic properties of our model. From this viewpoint an ideal embedding seems to be the most natural and simple type of embedding to study. Ideally embedded space-times receive the least amount of tension from the surrounding space. We found that ideally embedded hypersurfaces in a pseudo-Euclidean space contain the de Sitter spaces and a Robertson-Walker model. Embeddings of the de Sitter and Robertson-Walker models were already considered by Ponce de Leon\cite{ponce}. It was later realized that his 5-dimensional embedding space was flat\cite{wesson,mcmanus} and this was used in e.g. Ref.~\onlinecite{seahra2} to study the structure of the Big Bang. 

{\par\noindent}Furthermore a class of anisotropic perfect fluid models containing a timelike two-surface of constant curvature has also been shown to be ideally embedded. Because the non-flat vacuum models were excluded from our study due to Theorem~\ref{thszek} we will study them in a following paper.


\begin{acknowledgments}
The authors would like to thank the anonymous referee for pointing out references to earlier work.
\end{acknowledgments}


\end{document}